\documentclass[aps,apl,preprint,showpacs,groupedaddress]{revtex4}
\pdfoutput=1

\usepackage{dcolumn}   
\usepackage{bm}        
\usepackage{amssymb}   
\usepackage{verbatim}
\usepackage{graphicx}

\usepackage{amsmath}
\usepackage{amsfonts}
\usepackage{float}

\newcommand{\ket}[1]{|#1\rangle}

\def\lsim{\mathrel{\rlap{\lower4pt\hbox{\hskip1pt$\sim$}}
    \raise1pt\hbox{$<$}}}                
\def\gsim{\mathrel{\rlap{\lower4pt\hbox{\hskip1pt$\sim$}}
    \raise1pt\hbox{$>$}}}                

\begin{document}

\title {High Fidelity Qubit Readout with the Superconducting Low-Inductance Undulatory Galvanometer Microwave Amplifier}
\author{D. Hover}
\author{S. Zhu}
\author{T. Thorbeck}
\author{G.J. Ribeill}
\affiliation{Department of Physics, University of
Wisconsin, Madison, Wisconsin 53706, USA}
\author{D. Sank}
\author{J. Kelly}
\author{R. Barends}
\author{John M. Martinis}
\affiliation{Department of Physics, University of California, Santa Barbara, California 93106, USA}
\author{R. McDermott}
\affiliation{Department of Physics, University of
Wisconsin, Madison, Wisconsin 53706, USA}
\email[Electronic address: ]{rfmcdermott@wisc.edu}

\date{\today}

\begin{abstract}
We describe the high fidelity dispersive measurement of a superconducting qubit using a microwave amplifier based on the Superconducting Low-inductance Undulatory Galvanometer (SLUG). The SLUG preamplifier achieves gain of 19 dB and yields a signal-to-noise ratio improvement of 9 dB over a state-of-the-art HEMT amplifier. We demonstrate a separation fidelity of 99\% at 700 ns compared to 59\% with the HEMT alone. The SLUG displays a large dynamic range, with an input saturation power corresponding to 700 photons in the readout cavity.
\end{abstract}

\maketitle

Over the past decade, circuit quantum electrodynamics (cQED) has emerged as a powerful paradigm for scalable quantum information processing in the solid state \cite{Blais04, Wallraff04, Schuster05}. Here a superconducting qubit plays the role of an artificial atom, and a thin-film coplanar waveguide or bulk cavity resonator is used to realize a bosonic mode with strong coupling to the atom \cite{Majer07, Hofheinz08, Hofheinz09, DiCarlo09,DiCarlo10,Chow12}. In the limit where the qubit is far detuned from the cavity resonance, it is possible to perform a quantum nondemolition measurement of the qubit by monitoring the microwave transmission across the cavity at a frequency close to the cavity resonance \cite{Blais04}. In order to maximize measurement fidelity, it is necessary to reduce the added noise of the measurement system to the greatest extent possible \cite{Gambetta08}. For conventional HEMT-based readout at frequencies in the range from 6-7 GHz, one has an added system noise around twenty quanta, $n_a \sim 20$, and for typical parameters one finds single-shot qubit measurement fidelity of order 50\% in 500 ns. For fast, high fidelity, single-shot readout in cQED, it is necessary to read out the cavity with an amplifier whose noise performance approaches the standard quantum limit $n_a = 1/2$, the minimum noise achievable by a phase-insensitive linear amplifier \cite{Caves}.

There has been significant recent progress in the use of Josephson parametric amplifiers (JPAs) for qubit readout. Specific milestones include observation of quantum jumps in a transmon qubit \cite{Vijay11}, heralded state preparation to eliminate initialization errors \cite{Johnson12,Riste12}, and stabilization of qubit Rabi oscillations using quantum feedback \cite{Vijay12}. While JPAs achieve noise performance that approaches the standard quantum limit (or even surpasses it in the case of operation in phase-sensitive mode), state-of-the-art amplifiers still suffer from small instantaneous bandwidth and low dynamic range, with input saturation power of order -120 dBm \cite{Mutus13}. These features make it challenging to multiplex in a multi-qubit architecture, i.e., to read out multiple cavity tones simultaneously with a single JPA. Moreover, the JPA requires a separate strong microwave pump tone, from which the qubit must be protected by several stages of cryogenic isolation.
\begin{figure}
\centering
\includegraphics[height=67mm]{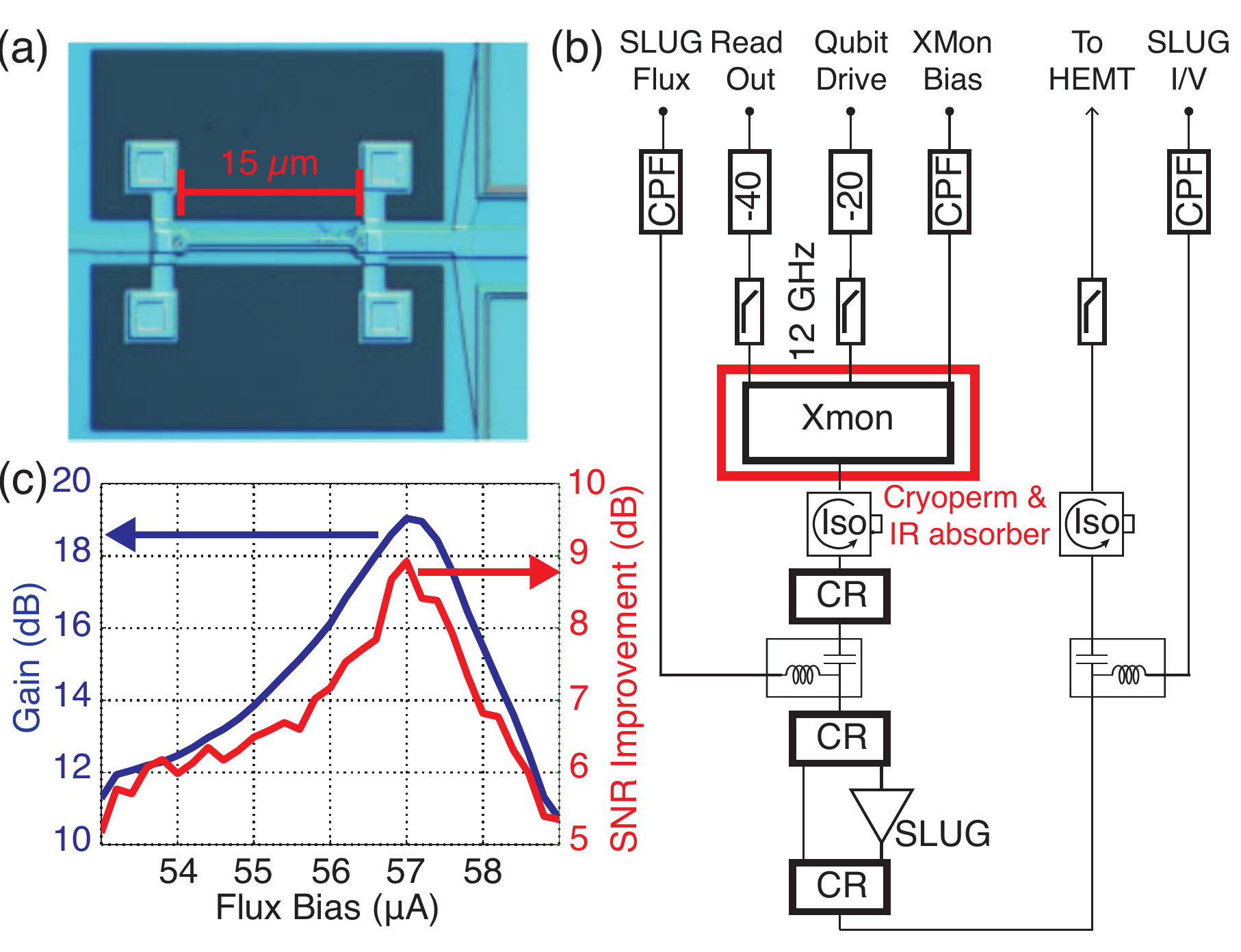}
\vspace*{-0.0in} \caption[]{(a) Micrograph of SLUG gain element. The $2~\mu$m$^2$ junctions are separated by a $15~\mu$m long SLUG body with trace width $1~\mu$m. The input signal is coupled to the SLUG \textit{via} a lumped element $LC$ matching network (not shown). (b) Block diagram of the experiment. All components shown here are mounted at the mixing chamber plate of the DR. Here CPF stands for copper powder filter, Iso. stands for cryogenic isolator, and CR stands for cryogenic coaxial relay. (c) Gain (blue) and SNR improvement (red) \textit{versus} flux bias of the SLUG amplifier close to the qubit readout frequency of 6.614 GHz. The device was biased at a current of $33~\mu$A. }
\label{fig:Wiring}
\end{figure}

In this Letter, we describe an alternative approach to qubit readout based on amplification with a Superconducting Low-inductance Undulatory Galvanometer (SLUG), a variant of the dc Superconducting QUantum Interference Device (dc SQUID) \cite{Ribeill11, Hover12}.  SQUID based amplifiers have demonstrated noise performance within a factor of two of the standard quantum limit at 600 MHz \cite{Clarke01} and large gain-bandwidth products at GHz frequencies \cite{Aumentado09}. Specific advantages of the SLUG geometry include a large flux-to-voltage transfer function, a relatively large real part of the input impedance of order a fraction of an ohm, and a compact device geometry that is free of parasitic stray reactances. While the noise of a SLUG amplifier will never equal that of an optimized JPA,  added noise of order one quantum is achievable at frequencies approaching 10 GHz. Moreover, an optimized SLUG has the possibility of achieving an instantaneous bandwidth approaching 1 GHz with saturation powers of -90 dBm \cite{Ribeill11}, opening the door to multiplexed single-shot readout in circuit QED.

A micrograph of the SLUG element is shown in Fig.~\ref{fig:Wiring}(a). The SLUG amplifier was realized in a six-layer process using optical projection lithography: three superconducting Nb layers with thickness 100 nm define the circuit groundplane and the two branches of the SLUG loop; two SiO$_2$ layers of the same thickness separate the superconducting traces; and a Pd layer with thickness 30 nm was used to realize the 8 $\Omega$ resistors shunting each junction. The Nb/Al-AlO$_x$-Al/Nb Josephson junctions were formed in $2~\mu$m$^2$ vias etched in the upper SiO$_2$ layer. The critical current per junction is $I_0 = 20~\mu$A, corresponding to a critical current density of 1 kA/cm$^2$. The mutual inductance between the input signal and the SLUG loop is $M=6.7$ pH, and the peak-to-peak voltage modulation of the device is around 130 $\mu$V. The input matching network is a single-pole lumped element $LC$ section with a designed characteristic impedance of $2~\Omega$; the component values were chosen to maximize the gain-bandwidth product at an operating frequency of 6.6~GHz. The 2x2 mm$^2$ die was packaged in a brass box and clamped onto the cold stage of a dilution refrigerator (DR).

The Xmon qubit has been described in detail elsewhere \cite{Barends13}. The device is capacitively coupled to a quarter-wave coplanar waveguide resonator with a measured coupling constant $g/2\pi = 30$ MHz,  a resonance frequency $\omega_0/2\pi$ near 6.6 GHz, and a cavity decay rate $\kappa/2\pi = 1.14$ MHz. The qubit frequency $\omega_{10}/2\pi$ is close to 6 GHz, with an anharmonicity of 230 MHz and a ratio $E_J/E_C \approx 95$ \cite{Koch07}. The qubit was mounted in a superconducting aluminum box inside a cryoperm shield. The use of nonmagnetic connectors and cabling in the vicinity of the sample ensured that the device was cooled in a low magnetic field environment, and multiple stages of infrared shielding suppressed quasiparticle generation from stray light \cite{Barends11}.  We measured a qubit energy relaxation time $T_1 =  10.4~\mu$s both with and without the SLUG in the amplification chain.

A block diagram of the experiment is shown in Fig.~\ref{fig:Wiring}(b). Three cryogenic coaxial relays at the mixing chamber of the DR allowed separate characterization of the qubit and the SLUG amplifier and enabled \textit{in situ} calibration of SLUG gain and signal-to-noise ratio (SNR) enhancement. The SLUG flux (current) bias was heavily filtered at both 4.2 K and 40 mK and combined with the microwave input (output) of the amplifier using a commercial bias-T. The output signal passed through a cryogenic isolator before being amplified by a low-noise HEMT at 4.2 K; a second isolator between the qubit and SLUG suppressed microwave emission from the SLUG input back toward the qubit. The measurement chain without the SLUG amplifier had a system noise of around twenty quanta, $n_a \sim 20$, at 6.6 GHz.

\begin{figure}
\centering
\includegraphics[height=67mm]{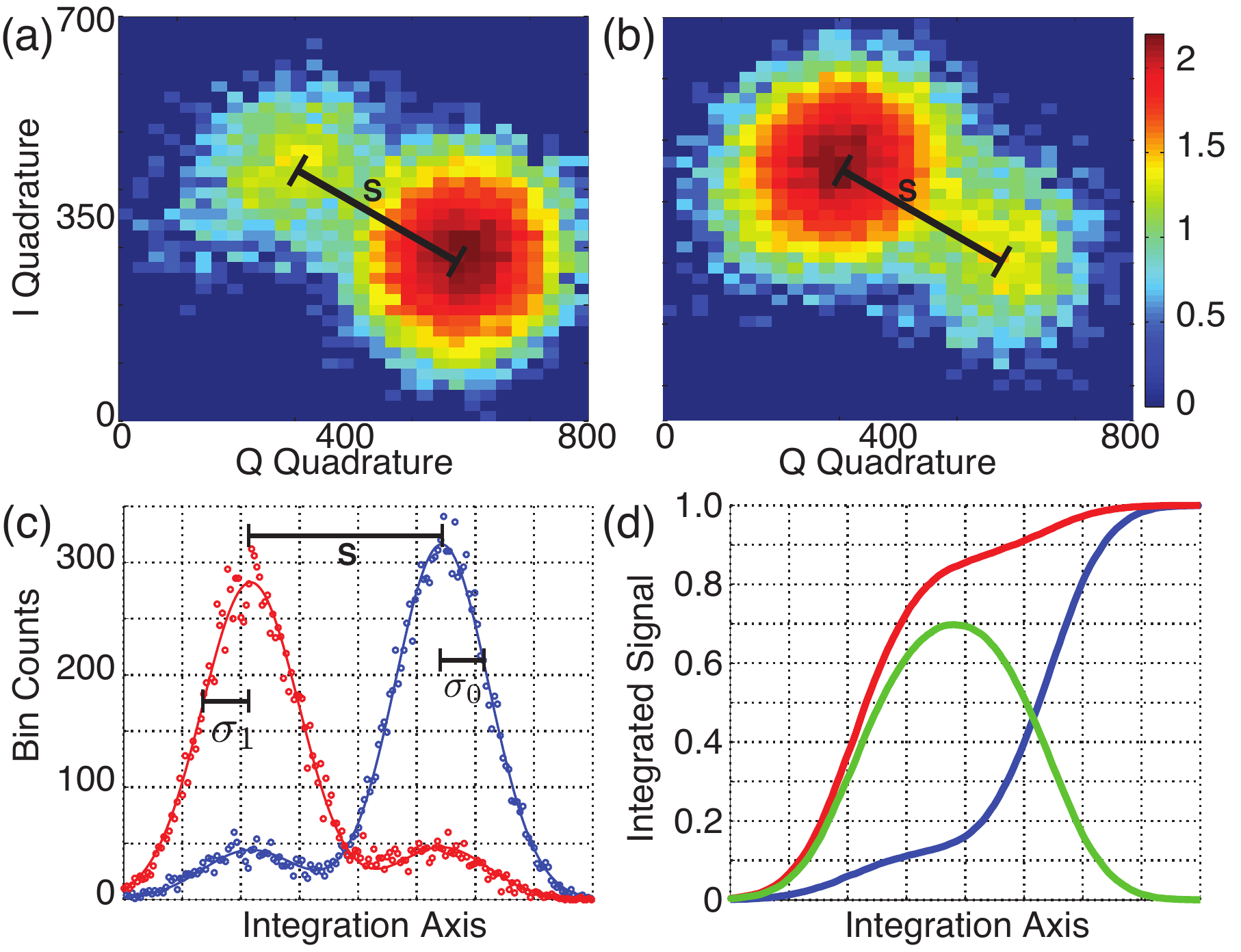}
\vspace*{-0.0in} \caption[]{Heterodyne signal distribution measured with the SLUG for qubit preparation in the $\ket{0}$ (a) and $\ket{1}$ (b) states. The measurement time was 600 ns and the readout power corresponds to a steady state photon occupation $\bar{n}=108$ in the readout resonator. The false color scale is the log of the bin counts. The optimal measurement axis is indicated by the black line, and $S$ is the distance between the centers of the two distributions in quadrature space.  Projected heterodyne histogram (c) and integrated signal (d) for qubit preparation in the $\ket{0}$ (blue) and $\ket{1}$ (red) state. The dashed lines in (c) are fits to a double Gaussian, while $\sigma_0$ and $\sigma_1$ are the standard deviations of the qubit state distributions. The integrated signals in (d) are normalized to the total number of qubit preparations, and the measurement fidelity (green) is the difference between the two integrated signals.}
\label{fig:IQFids}
\end{figure}

In Fig.~\ref{fig:Wiring}(c) we plot the measured gain and SNR improvement \textit{versus} SLUG flux bias at a frequency slightly detuned from the qubit measurement frequency. At optimal bias, the SLUG achieved 19 dB gain while improving the SNR of the amplification chain by 9 dB. The instantaneous bandwidth of this SLUG device was around 50 MHz at the qubit measurement frequency and could be dynamically tuned from 6.45 to 6.65 GHz.

\begin{figure}
\centering
\includegraphics[height=44mm]{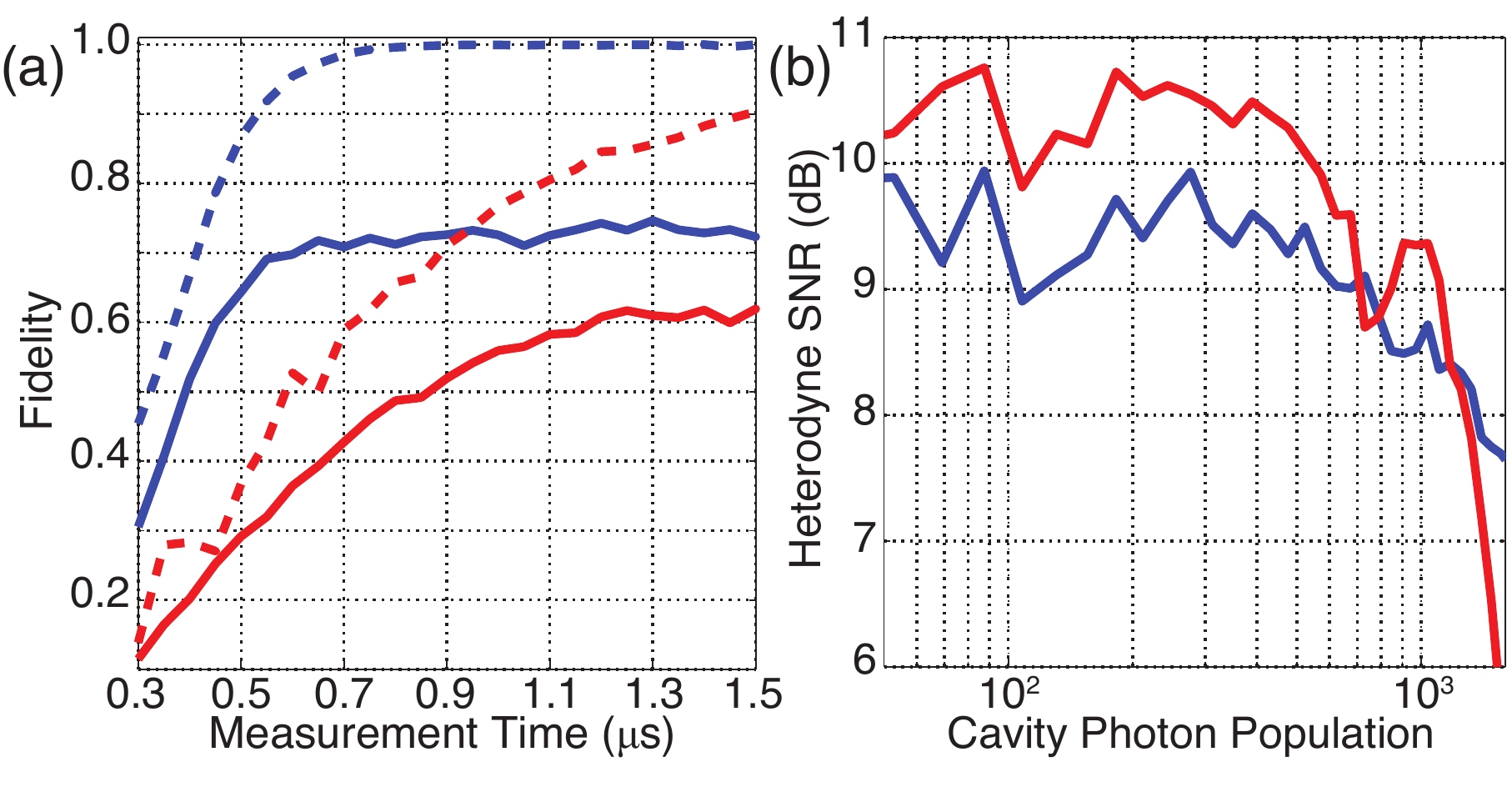}
\vspace*{-0.0in} \caption[]{(a) Qubit measurement fidelity obtained with (blue) and without (red) SLUG preamplification. The cavity drive corresponds to a steady state photon occupation $\bar{n} = 108$. The solid (dashed) lines are the raw measurement fidelities (separation fidelities); see main text. (b) Heterodyne SNR improvement with the SLUG \textit{versus} cavity photon occupation for measurement times of 600 ns (blue) and $1.2~\mu$s (red). }
\label{fig:Fidelities}
\end{figure}

The average photon occupation of the cavity was calibrated with a measurement of the ac Stark shift of the qubit frequency \cite{Schuster05, Boissonneault10}. Here the cavity was excited by the readout tone for a duration much longer than its characteristic time $\tau_c = 2\pi/\kappa = $ 880 ns, ensuring that the dynamics of the cavity had reached their steady state before the qubit frequency was probed spectroscopically.  As a result, the average photon occupation $\bar{n}$ reported in this Letter overestimates the actual occupation for measurement times on the order of $\tau_c$.

To measure the qubit, we monitored the microwave transmission across the cavity at a frequency of 6.614~GHz, corresponding to the dressed cavity resonance observed when the qubit is in the $\ket{0}$ state. For each measurement shot, the heterodyne voltage was integrated over time and the in-phase (I) and quadrature (Q) components of the signal were recorded. In this measurement scheme, multiple preparations of the $\ket{0}$ or $\ket{1}$ states yield two different Gaussian distributions in IQ space. In Figs.~\ref{fig:IQFids}(a-b) we plot the heterodyne signal distributions for nominal preparation of the $\ket{0}$ and $\ket{1}$ states, respectively. In each of these plots, we observe one main Gaussian component corresponding to the desired state, with a small satellite component corresponding to a spurious admixture of the other qubit state. The separation of the heterodyne distributions corresponding to the qubit basis states is determined by the amplitude and duration of the microwave drive and by the state-dependent cavity pull, while the widths of the distributions reflect the overall noise of the measurement chain. 
The line joining the centers of the $\ket{0}$ and $\ket{1}$ state distributions defines an optimal measurement axis for heterodyne detection. In Fig.~\ref{fig:IQFids}(c) we plot the heterodyne histograms projected against this axis for
qubit preparation in the $\ket{0}$ (blue) and $\ket{1}$ (red) states; here $\ket{0}$ state preparation was accomplished by waiting many qubit $T_1$ times prior to measurement, while the $\ket{1}$ state was prepared from the $\ket{0}$ state with a calibrated $\pi$ pulse. In these experiments, the readout was driven with an amplitude corresponding to a steady-state $\bar{n} = 108 \approx n_{crit}$, where the critical photon number $n_{crit} \equiv \Delta^2/4g^2$  roughly marks the breakdown of the dispersive approximation to the Jaynes-Cummings Hamiltonian \cite{Blais04, Gambetta07}.

We evaluate measurement fidelity by integrating the two histograms along the optimal measurement axis and taking the difference, as shown in Fig.~\ref{fig:IQFids}(d). Measurement fidelities for SLUG (blue) and HEMT (red) are plotted as solid traces in Fig.~\ref{fig:Fidelities}(a) for a range of measurement times $\tau_m$, where $\tau_m$ is the duration of the cavity excitation pulse. The measurement fidelity plateaus at 0.7 for $\tau_m = 700$ ns when using the SLUG amplifier; this represents a significant improvement compared to the measurement fidelity of 0.3 with the HEMT alone for the same measurement time. Several factors contribute to measurement infidelity: (1) state preparation errors, (2) relaxation errors, (3) measurement-induced transitions, and (4) measurement errors. Errors (1)-(3) are the result of a transfer of weight from one of the two heterodyne state distributions to the other, while measurement errors are the result of insufficient separation of the distributions in quadrature space due to excessive noise of the measurement system. For this reason, it is possible to separate out preparation/relaxation/induced transition errors from measurement errors by fitting two Gaussian components to the heterodyne data. For example, preparation errors lead to a large spurious second Gaussian component in the quadrature histograms. To focus on the effect of the measurement chain alone, one can ignore the spurious second component and rescale the desired component appropriately in order to evaluate a separation fidelity, a figure of merit that is independent of qubit preparation and relaxation errors. In our experiments, measurement errors are dominant at short times (under $\sim$400 ns), while for longer times we find that state preparation errors dominate. Indeed, we observe a large equilibrium $|1\rangle$ state population around 12\% when we are nominally initializing in $\ket{0}$; we believe that this is due to insufficient filtering of IR radiation. This preparation error, which is present with and without the SLUG in the amplification chain, contributes twice to measurement infidelity, degrading fidelity of both the $\ket{0}$ state and the $\ket{1}$ state, which is prepared from the corrupted $\ket{0}$ state. Thus, our excess $\ket{1}$ state population of 12\% immediately limits measurement fidelity to 76\%. From the measured $T_1$ time of 10.4 $\mu$s we expect an infidelity of 2.5\% in 500 ns.

We display the separation fidelities of qubit measurements with and without the SLUG preamplifier as dashed traces in Fig.~\ref{fig:Fidelities}(a). We find a separation fidelity of 0.99 with the SLUG compared to 0.59 without the SLUG for a measurement time of 700 ns. This improvement is due entirely to the significant reduction in added noise of the measurement chain at the qubit measurement frequency. We remark that the measurement time is limited by the rate at which photons leak out of the readout cavity, which in the current experiment is relatively weakly coupled to the measurement apparatus. We anticipate that the incorporation of an appropriate Purcell filter at the output of the cavity will enable strong coupling at the cavity output without introducing additional dissipation at the qubit frequency \cite{Reed2010}, so that measurement fidelities of order 99\% should be attainable in times of order 100 ns.

One of the advantages of the SLUG amplifier compared to the JPA is the significantly higher saturation power. In Figure~\ref{fig:Fidelities}(b), we plot the heterodyne SNR improvement as a function of number of photons in the resonator for measurement times of 600 ns and $1.2~\mu$s; here, we define heterodyne SNR as the ratio $S/(\sigma_0+\sigma_1)$ of the separation $S$ of the centers of the $\ket{0}$ and $\ket{1}$ histograms to the sum of their standard deviations. Here we extrapolated $\bar{n}$ from low power by assuming a linear relationship  between the power applied to the chip and the cavity occupation.  The 1 dB compression point occurs at $\bar{n}$ between 700 and 1000 photons in the cavity, an average cavity photon occupation number that exceeds the saturation power of a typical JPA by almost an order of magnitude \cite{Lin13}.

In conclusion, we have described high fidelity single-shot measurements of an Xmon qubit using an ultralow-noise SLUG microwave amplifier. The SLUG improves the SNR of the measurement chain by around 9 dB, yielding an improvement in single-shot separation fidelity from 0.59 to 0.99 in 700 ns. The measurement time is limited by the rate at which photons leak out of the readout cavity; with the addition of an appropriate Purcell filter, we anticipate that similar fidelities could be reached in measurement times of order 100 ns. With the potential for large instantaneous bandwidth and a saturation power that is more than an order of magnitude greater than that of an optimized JPA, the SLUG is well suited to frequency-multiplexed dispersive readout of multiple superconducting qubits, for example in a scalable surface code circuit \cite{Fowler12}.

This research was funded by the Office of the Director of National Intelligence (ODNI), Intelligence Advanced Research Projects Activity (IARPA), through the Army Research Office Grants W911NF-10-1-0334 and W911NF-11-1-0029. All statements of fact, opinion, or conclusions contained herein are those of the authors and should not be construed as representing the official views or policies of IARPA, the ODNI, or the U.S. Government.


\end{document}